\documentstyle[twocolumn,epsbox]{jpsj}
\def\runtitle{
Luttinger-liquid Parameter of  Hubbard Chain  and Hubbard Ladder 
}
\def\runauthor{
Kazuhiro {\sc Sano}   
}

\newcommand{\simk}{\stackrel{<}{_\sim}}

\title
{
Luttinger-liquid Parameter of  Hubbard Chain  and Hubbard Ladder 
}
\author
{ 
Kazuhiro {\sc Sano}  
}
\inst
{
 Department of Physics Engineering,  Mie University, Tsu, Mie 514-8507, Japan 
}
\date{\today}
\abst{
 We study the Luttinger-liquid parameter  $K_{\rho}$ of the  Hubbard chain  and the Hubbard ladder models by the ordinary perturbation method  combined with  the Luttinger-liquid  relation. 
According to the Luttinger-liquid relation,
the critical exponent $K_{\rho}$ is related to the charge susceptibility  $\chi_c$ and the Drude weight $D$ by  $  K\sb{\rho}=\frac{1}{2}(\pi \chi_c D)^{1/2}$.
By calculating these quantities with the perturbation method,   we  obtain  $K_{\rho}$ at   the first-order   analytically and up  to the second-order numerically. 
We  compare these results  with results of the  Bethe ansatz  for the  Hubbard chain and that of the numerical diagonalization  for the Hubbard ladder.
It shows that   the perturbation calculation of $K_{\rho}$ is  reliable in the weak coupling regime. 
}
\kword
{
Luttinger-liquid, perturbation, Hubbard Ladder,superconductivity
}
\begin{document}
\sloppy
\maketitle
\section{Introduction}
 Low-dimensional strongly  correlated electron systems   have been investigated extensively due to the possible relevance to high-$T_c$ superconductivity.
In particular, One-dimensional(1D) Hubbard-like models provide an important  ground for understanding electron-correlation effects.
It is well known that  a single Hubbard chain has generic properties of Luttinger liquid state with gapless spin and charge modes for repulsive interaction. For attractive interaction, a superconducting state  characterized by dominant paring correlations with a spin gap is realized. 
\cite{Solyom,Haldane,Voit}

Recently, weak coupling theory  has been applied to the problem of ladder models, which are interesting as a first step towards 2D systems and may be relevant for some materials.
 The theory reveals that the systems remain typical non-Fermi-liquid properties as   1D electronic systems.\cite{F,S,B}
At half-filling, they are Mott insulators, exhibiting gaps to all excitations. Upon doping, the gaps survive except one gapless charge mode and a  superconducting (SC) paring correlation characterized slowly power-low decay appears.
According to the Luttinger-liquid theory,  critical exponents of various types of correlation functions are determined by a single  parameter $K_\rho$.\cite{Solyom,Haldane,Voit}
 It is  predicted that 
 the SC correlation function of  ladder systems is dominant for $K_{\rho}>0.5$ at lightly doping. 
It decays   as $\sim r^{-(\frac{1}{2K_{\rho}})}$,  whereas 
 the "$4k_F$"  charge density wave (CDW)  correlation  function decays  as $\sim r^{-2K_{\rho}}$. The spin density wave (SDW)  and "$2k_F$"  CDW correlation functions decay exponentially.\cite{F,S,B}

In spite of the good understanding of  the single Hubbard chain
which is exactly solved by the Bethe ansatz method,   Hubbard ladder models are much less known. 
  In fact,  systematic  treatment of $K_\rho$  as a function of  interaction $U$ is not  yet obtained.
In this work we propose a simple method for calculating  the Luttinger-liquid parameter  $K_{\rho}$. 
It is  the  first and second-order perturbation expansion with respect to  $U$  combined with  the Luttinger-liquid  relation. 
These results are compared with the  Bethe ansatz  and  the numerical diagonalization results.

\section{  Perturbation expansion of  $K_{\rho}$ }

In the Luttinger liquid theory,  some relations  have been  established as universal relations in one-dimensional  models.\cite{Solyom,Haldane,Voit}  
The critical exponent $K_{\rho}$ is related to  the charge susceptibility $\chi$ and  the Drude weight $D$ by
\begin{equation}
      K\sb{\rho}=\frac{1}{2}(\pi \chi D)^{1/2},
\end{equation}
with 
\begin{equation}
\chi^{-1}= \frac{1}{N}\frac{\partial^2 E_g(n)}{\partial n^2},
\qquad
D=\frac{\pi}{N} \frac{\partial^2 E_g(\phi)}{\partial \phi^2},
\end{equation}
 where $N$ is the number of lattice site, $n$ is electron density
 and 
 $E_g$ is the total energy of the ground state as a function of $n$ and a magnetic flux $\phi$.\cite{Voit}
If we  calculate  the ground state energy in some way, 
 we can  obtain   $K_{\rho}$ through 
 $\chi$ and   $D$. 

 We apply  the ordinary perturbation expansion in powers of $U$ to $ E_g$, 
$$ E_g=E_0+e_1+e_2+......, $$
where $E_0$  is  the  ground state energy of the non-interacting system and $e_i$ ($i$=1,2,....) is the $i$-th order correction of $E_g$.
According to the perturbation expansion of $E_g$, 
 $\chi^{-1}$ and  $D$ are given by 
$$\chi^{-1}=\chi_0^{-1}+x_1U+x_2U^2....,\quad D=D_0+d_1U+d_2U^2+....,$$
where $\chi_0^{-1}$ and $D_0(=4\chi_0^{-1}/\pi)$ are 
 the inverse charge susceptibility and  the Drude weight
of the non-interacting system respectively. 
Coefficients $x_i$ and $d_i$ ($i$=1,2,....) are the $i$-th order corrections of $\chi^{-1}$ and $D$, which are determined by
the   second  differential coefficients of $e_i$ with respect to $n$ and $\phi$ respectively.
In the Hubbard model, the first-order term $e_1$ is easily obtained as $e_1=U\sum_{i}<n_{i\uparrow}><n_{i\downarrow}>=UNn^2/4$.
It leads  $x_1=1/2$ and $d_1=0$,  where $e_1$   is independent of the flux $\phi$. 

Substituting these values into eq. (1), we  obtain  $K_{\rho}$ within the first perturbation expansion:
\begin{equation}
      1/K\sb{\rho}^2 \simeq 1+\frac{\chi_0}{2}U.
\end{equation}
It shows that  $K_{\rho}$ generally decreases with increasing the repulsive interaction $U$ in the weak coupling limit.     
It is noted that this approximation 
is  equivalent to the Hartree-Fock (HF) approximation.
The second-order term of the ground state energy  $e_2$
is also easily obtained as
\begin{eqnarray} 
  e_2=\frac{U^2}{N}\sum_{k_1,k_2,q,\sigma}\frac{f(k_1)f(k_2)(1-f(k_1+q))(1-f(k_2-q))}
  {\varepsilon(k_1)+\varepsilon(k_2)-\varepsilon(k_1+q)+\varepsilon(k_2-q)},     \nonumber 
 \end{eqnarray} 
where $\varepsilon(k)$ is the non-interacting band and $f(k)=\theta(k_F-|k|)$.
Using $e_2$, we can calculate $K_{\rho}$  up to order $U^2$ by
\begin{eqnarray} 
1/K_{\rho}^2\simeq 1+\frac{\chi_0}{2}U+(x_2-\frac{\pi}{4}d_2)\chi_0U^2,
\end{eqnarray} 
with
$$x_2= \frac{1}{NU^2}\frac{\partial^2 e_{2}}{\partial n^2}, \qquad
d_2=\frac{\pi}{NU^2} \frac{\partial^2 e_2}{\partial \phi^2}.$$
In the following section, we will estimate the above integral  and obtain  $x_2$ and $d_2$  numerically.

\begin{figure}
\vspace{0.8cm}
\psbox[height=7.5cm]{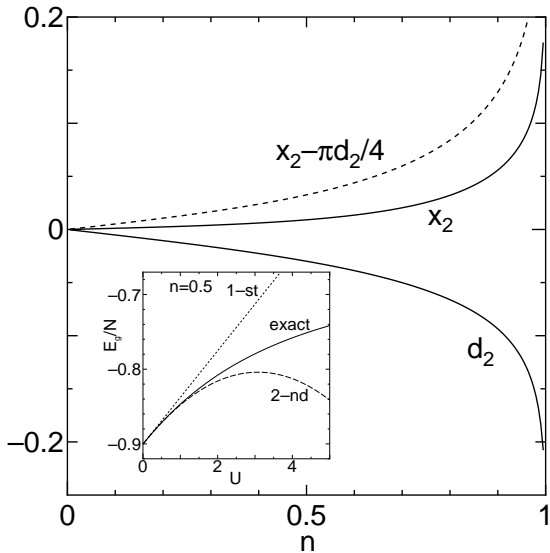}
%
\caption{
The coefficients of the second-order terms $x_2$ and $d_2$ as a function of $n$. 
The broken line represents  $(x_2-\frac{\pi}{4}d_2)$.
The inset shows the ground state energy of the system as a function of $U$ by the first-order and the second-order   perturbation calculation with the exact result at  quarter-filling.
}
\label{fig:1}
\end{figure}
\section{  Hubbard chain and  Hubbard ladder  }
At first, we  examine  the  Hubbard chain model
whose non-interacting band is given by $\varepsilon(k)=-2t\cos k$. 
In this case, we have  $\chi_0^{-1}=\pi t \sin k_F$ and $K_{\rho}$ up to the first-order expansion of $U$ 
\begin{equation}
      1/K\sb{\rho}^2=1+U/2\pi t \sin k_F,
\end{equation}
where   $k_F=\frac{\pi }{2}n$ is the Fermi wave number.
This expression of $K_{\rho}$  is exactly equal to  the result of bosonization method for the  Hubbard chain.\cite{Solyom}

To estimate $e_2$, we use   $N=200$ and 400 sites systems.
  We confirm that the size dependence  of  $x_2$ and $d_2$  are very small and negligible.
 In Fig.1, we show $x_2$ and $d_2$ as a function of $n$.
These values seem to diverge at the limit $n\rightarrow 1 $. 
It might reflect the insulator transition of the Hubbard chain at   half-filling.
We also show
 the ground state energy $E_g/N$ as a function of $U$ at quarter-filling. 
The result of $E_g/N$ is consistent with the exact result in the weak coupling region. 
\begin{figure}
\vspace{0.8cm}
\psbox[height=6.5cm]{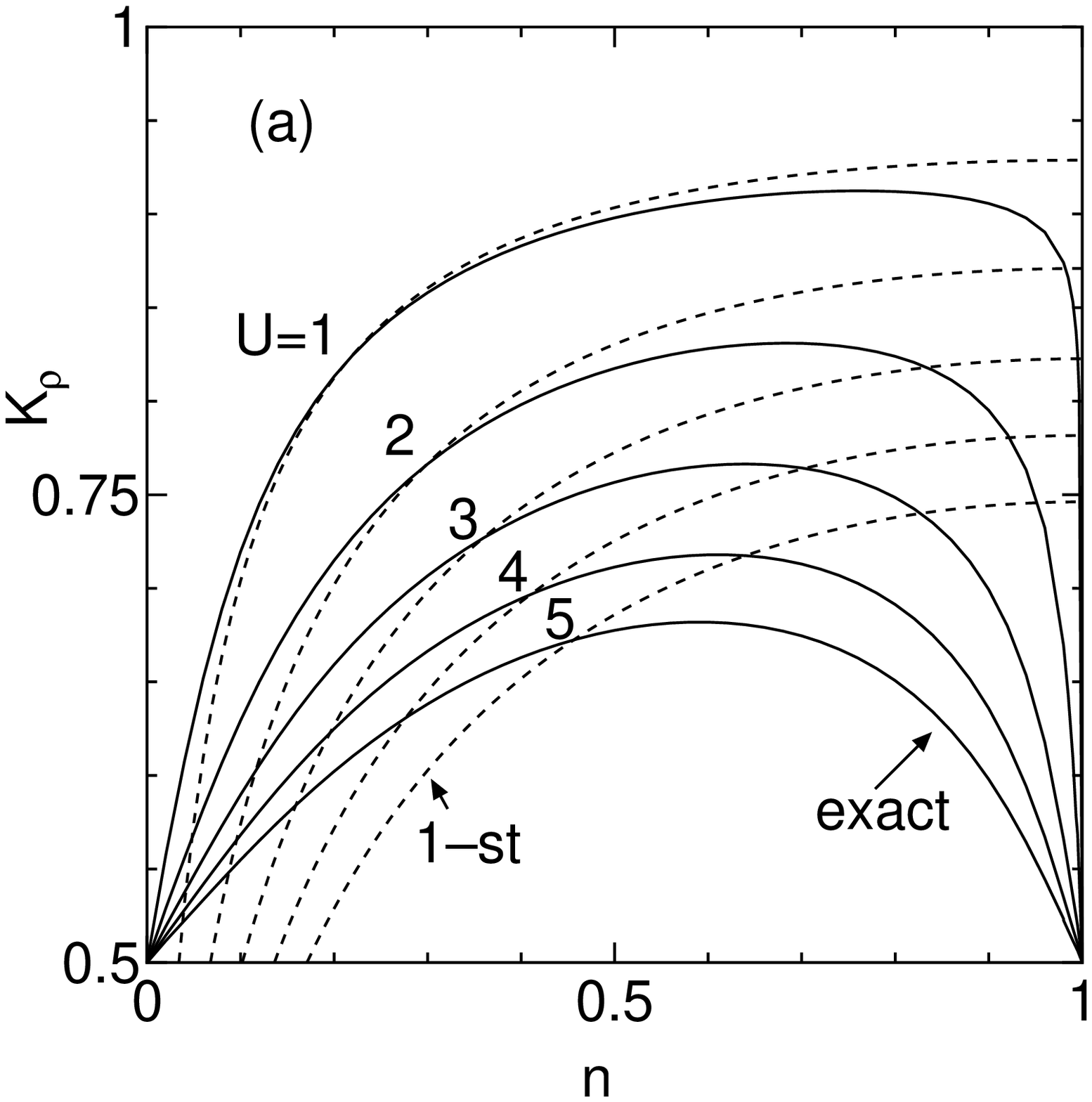}
\end{figure}
\begin{figure}
\vspace{0.8cm}
\psbox[height=6.5cm]{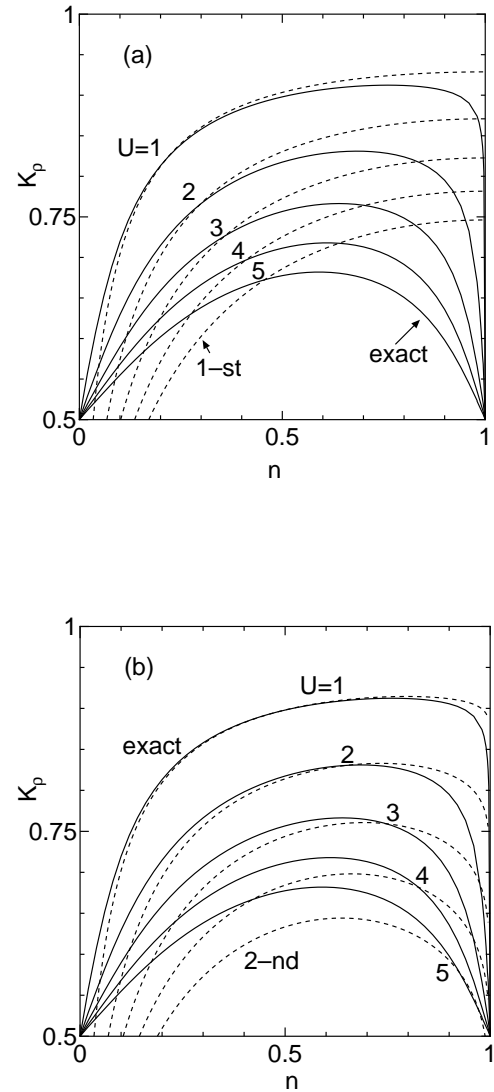}
%
\caption{
$K_{\rho}$ as a function of $n$ by (a) the first-order  and (b) the second-order perturbation expansion (b) with the exact result of the Bethe ansatz.
The  solid lines represent the results of the Bethe ansatz and the broken lines are that of the perturbation expansion.
}
\label{fig:2b}
\end{figure}
In Figs.2(a) and (b), we show  $K_{\rho}$ by the perturbation expansion  with the exact result of the Bethe ansatz.\cite{Lieb,Schulz}
They show that  the result of the perturbation expansion is consistent with the exact result in the weak coupling region. 
It also indicates that  the  second-order  perturbation calculation agrees with  the exact solution more than  the first-order calculation  in the weak coupling regime.
In the strong coupling regime, $K_{\rho}$ of the first-order calculation  seems to be close that of the exact solution  at $n\sim 0.4$.
However, it  may be  an accidental.

Next, we consider the Hubbard ladder Hamiltonian 

\begin{eqnarray} 
  H&=&-t_l\sum_{i,\alpha,\sigma} c_{i,\alpha,\sigma}^{\dagger} c_{i+1,\alpha\sigma}
   -t_r\sum_{i,\sigma} c_{i,1,\sigma}^{\dagger} c_{i,2,\sigma}+h.c. \nonumber \\
    &+&U\sum_{i,\alpha}n_{i,\alpha \uparrow}n_{i,\alpha \downarrow},
\end{eqnarray} 
where $c^{\dagger}_{i,\alpha,\sigma}$  stands for a creation operator of an electron with spin $\sigma$  at site $(i,\alpha)$ and $U$ is the on-site  interaction.
 Here,    $\alpha(=1,2)$ denotes legs  and $i$ is rung. 
In this case,  non-interacting band is written as
$$
     \varepsilon^{\pm}(k)=-2t_{l}\cos k \pm t_{r}, 
$$
where $\varepsilon^{+}(k)(\varepsilon^{-}(k))$ represents the upper (lower) band and $k$ is the wave vector.
 If we define $k_{F-}$($k_{F+}$) as a Fermi point in the lower (upper) band, we find that $\varepsilon^{-}(k_{F-})=\varepsilon^{+}(k_{F+})$ with $k_{F-}+k_{F+}=k_F=n\pi$.
The  differential coefficient of $\varepsilon^{-}(k_{F-})$ with respect to $n$ leads the inverse charge susceptibility $\chi^{-1}_0$.
After a bit of calculation, we get 
\begin{eqnarray} 
\chi^{-1}_0&=&-t_{l}\pi \sin k_F
    \{ \frac{ \cos k_F(1-\cos k_F)+(t_r/t_l)^2-\sin^2 k_F} {( 1-\cos k_F)^2} \}
\nonumber \\
    &\times&  \{ (t_r/t_l)^2-2\frac{(t_r/t_l)^2- \sin^2 k_F }{ 1- \cos k_F } \}^{-1/2} .\nonumber
\end{eqnarray} 
Substituting $\chi^{-1}_0$ to eq.(3), we have $K_{\rho}$ of the Hubbard ladder model analytically within the first-order perturbation method.\cite{cmode}

When $n$ is smaller than $n_c$ which is determined by $\varepsilon^{+}(0)=\varepsilon^{-}(\frac{\pi n_c}{2})$, electrons are filled only in the lower band.     
The density of state of the lower band   is a half of that of the chain model.
Then,  $\chi_0$  is  given by $1/(2\pi\sin k_F)$ which is a half of that of the chain model.  
Within the first-order perturbation calculation,   the upper band is irrelevant and the correction of $K_{\rho}$  becomes a half of that of  the chain model.

As well as the Hubbard chain  model, we estimate   the   second-order terms  $e_2$, $x_2$ and $d_2$ numerically.\cite{ND}
 In Fig.3, we show $x_2$, $d_2$ and $x_2-\frac{\pi}{4}d_2$  as a function of $n$ at $t_r/t_l=1.0$.
For $n>n_c=0.5$,  the values of $x_2$ seem to be  small, but the absolute values of $d_2$ are large.
It shows that the values $x_2-\frac{\pi}{4}d_2$ are  positive and large.
It indicates that the second-order term of the repulsion $U$ reduces $K_{\rho}$  as well as the first-order term.
This result   does not contradict  the behavior of the SC correlation  function obtained by  density matrix renormalization group  and  quantum Monte Carlo methods.\cite{DMRG,QMC,DMRG2}    
Although system sizes used in these numerical works are too small to determine $K_{\rho}$ precisely, we find that a rough estimate indicates  $0.5\simk K_{\rho} \simk 1.0$.

On the other hand, for $n<0.5$,  the values $x_2-\frac{\pi}{4}d_2$ are negative.
In particular, the absolute values  are very large near $n=0.5$.
It indicates that $K_{\rho}$ is  enhanced by the second-order term.
In contrast to the case for $n>0.5$, the existence of the upper band 
seems to produce an effective attraction in the second-order correction.
 This result may consist with an enhancement of the SC correlation shown by the numerical diagonalization method.\cite{yamaji}
Figure 4 indicates  $K_{\rho}$ as a function of $n$ at $t_l/t_r=1.0$ by the first- and the second-order perturbation calculation. 
\begin{figure}
\vspace{0.8cm}
\psbox[height=6.5cm]{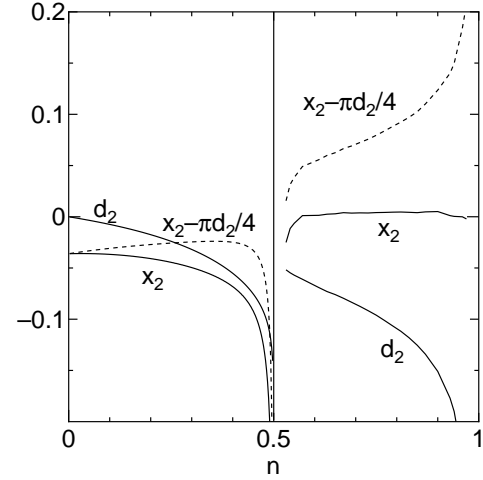}
%
\caption{
The coefficients  $x_2$ and $d_2$ as a function of $n$ for the Hubbard ladder model. 
The broken lines represent  $(x_2-\frac{\pi}{4}d_2)$.
}
\label{fig:3}
\end{figure}
\begin{figure}
\vspace{0.8cm}
\psbox[height=6.5cm]{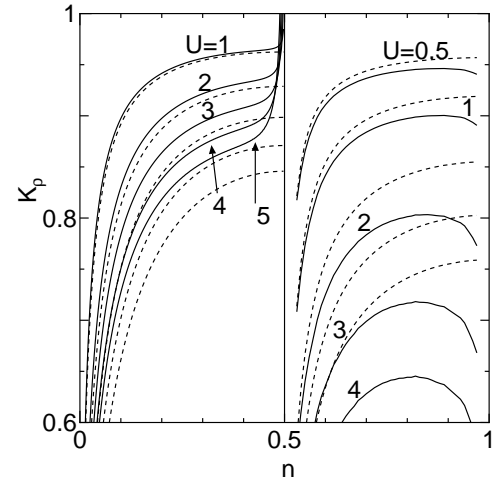}
%
\caption{
$K_{\rho}$ as a function of $n$ for the Hubbard ladder model by the first-order (the  broken lines)  and the second-order (the solid lines) perturbation expansion.
}
\label{fig:4}
\end{figure}

To clarify the validity of  the  perturbation expansion of $K_{\rho}$, we  examine   a finite size system of  the ladder model.
 We numerically diagonalize the Hamiltonian  of 14 sites (7 unit  cells) 
 system  by using the  Lanczos algorithm. 
We use the periodic boundary  condition for $N_e=8$, the  Moebius boundary  condition for $N_e=10$ and the antiperiodic boundary condition for $N_e=12$,  where $N_e$ is the total electron number.\cite{Muller}
 This choice of the boundary condition   gives either fully occupied or empty single particle orbitals ( closed shell )   
and   removes accidental degeneracy in the non-interacting case.
The uniform  charge susceptibility $\chi_c$ and the Drude  weight $D$ is calculated from the ground state energy with the usual method.\cite{Sano1} 
\begin{figure}
\vspace{0.8cm}
\psbox[height=7.5cm]{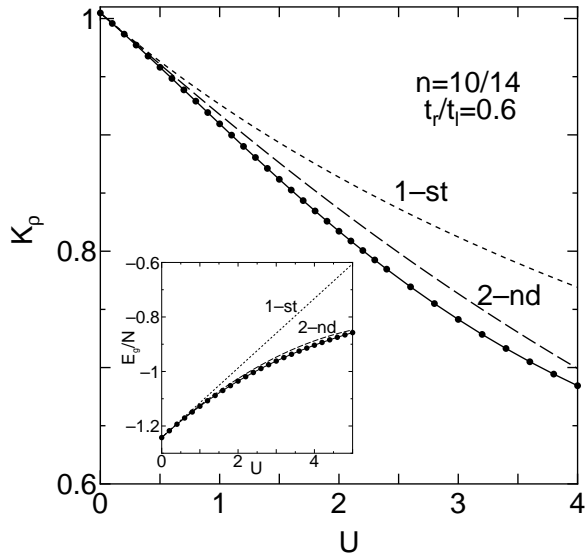}
%
\caption{
 $K_{\rho}$   of the  finite size system for the  Hubbard ladder as a function of $U$.  The solid circles represent the result of the numerical diagonalization for 7 rungs system at $t_r/t_l=0.6$. 
The  broken line represents the result of the first-order   perturbation expansion and the dashed line is that of the second-order   perturbation expansion.
The inset shows the ground state energy of the system as a function of $U$ by the first-order and the second-order   perturbation calculations with the result of the numerical diagonalization.
}
\label{fig:5}
\end{figure}

In Fig.5, we show  $K_{\rho}$ and the ground state energy $E_g/N$ of the  finite size  system with the result of the numerical diagonalization method.
In the weak coupling regime, the results of the perturbation approximation and  the numerical diagonalization  are in agreement with  each other. 
It also shows that  the second-order perturbation  is better than the first-order  in the weak coupling regime. 

\section{ Summary and discussion }
In this work we examine  the Luttinger-liquid parameter  $K_{\rho}$ of the  Hubbard chain  and the Hubbard ladder models by the ordinary perturbation method combined with  the Luttinger-liquid  relation. 
According to the Luttinger-liquid relation,
we obtain  $K_{\rho}$ at   the first-order   analytically and up  to the second-order numerically.
Comparing   $K_{\rho}$  with  the exact result of Bethe ansatz and that of the numerical diagonalization method, we  show that  the analysis of perturbation method is  reliable in the weak coupling region. 
 
Generally speaking, the validity of  the  perturbation expansion is not always obvious in 1D electron systems.\cite{shiba}
However, it has been analytically shown that  the ordinary  perturbation expansion of the ground state energy of the Hubbard chain agrees with  the expansion of the exact Bethe ansatz solution  at half-filling.\cite{Takahasi,Economou,Metzner}
Although the convergence radius is zero, the perturbation  is possible as  an asymptotic expansion.

Away from half-filling,  the integral equations of the  Bethe ansatz  solution are not analytically solved.
However, we confirm that $e_2$  consists with the second-order term of $U$  in the  Bethe ansatz  solution numerically. 
It suggests that  the  perturbation expansion  in powers of $U$ is also possible as an asymptotic expansion.\cite{asympo}


\begin{thebibliography}{99}
\bibitem{cmode}
It is noted that application of our analysis is limited to  systems which have  one gapless charge mode at the low  energy limit.
The number of the gapless charge mode  is not always  one in the ladder model. See ref. [4].
\bibitem{ND}
For $n>n_c$,  finite size effect of $x_2$ is large because Fermi points exist  in both of the upper and the lower bands.
To reduce it, we fit $e_2$ to a polynomial of  $n$ with various density  points  and estimate $x_2$ by  the second-order coefficient of the polynomial.
\bibitem{asympo}
If the convergence radius of the perturbation expansion is not zero, the electronic state of the system should be not  {\it "Luttinger-liquid" } but  {\it "Fermi-liquid" } in the weak coupling limit.  

\vspace{5mm}
\end{thebibliography}
\end{document}